# UV Direct-Writing of Metals on Polyimide

Jack Hoyd-Gigg Ng[1,2], Marc P.Y. Desmulliez[1,2], Aongus McCarthy[2], Himanshu Suyal[2], Kevin A. Prior[2] and Duncan P. Hand[2]

[1]Microsystems Engineering Centre (MISEC),
[2]School of Engineering and Physical Sciences,
Heriot-Watt University, Edinburgh EH14 4AS, Scotland, United Kingdom
E-mail: jhn4@hw.ac.uk

*Abstract* Conductive micro-patterned copper tracks were fabricated by UV direct-writing of a nanoparticle silver seed layer followed by selective electroless copper deposition. Silver ions were first incorporated into a hydrolyzed polyimide surface layer by wet chemical treatment. A photoreactive polymer coating, methoxy poly(ethylene glycol) (MPEG) was coated on top of the substrate prior to UV irradiation. Electrons released through the interaction between the MPEG molecules and UV photons allowed the reduction of the silver ions across the MPEG/doped polyimide interface. The resultant silver seed layer has a cluster morphology which is suitable for the initiation of electroless plating. Initial results showed that the deposited copper tracks were in good agreement with the track width on the photomask and laser direct-writing can also fabricate smaller line width metal tracks with good accuracy. The facile fabrication presented here can be carried out in air, at atmospheric pressure, and on contoured surfaces.

## I. INTRODUCTION

Direct-writing techniques of metallic patterns, such as laser or ink-jet printing, are additive processes which eliminate the traditional resist moulding and mask-based photo-patterning steps. Such techniques reduce the amount of wasted materials, the turnaround time and manpower cost. Laser, electron-beam and X-ray direct-writing have been employed for a long time in an enclosed pressurized chamber [1]. In search of low cost direct-writing processes without using any vacuum facilities, new types of processes have emerged which can be carried out through the coating of liquid solution or solid over-layer. The chart in Fig. I shows a review of the recent approaches employed by researchers worldwide. To minimize the effect of diffraction of light through liquid, direct-writing through a thin solid over-layer coating on the substrate is favored in order to achieve high resolution of sub-micron features. We propose a novel technique of direct laser writing, at atmospheric pressure, of metallic patterns onto a polyimide.

Polyimide is a high performance polymer widely used in microsystems technology owing to its many desirable properties such as low dielectric constant (low-k), high mechanical strength, high resistance to moisture and heat ($T_g$ = 365°C), chemical inactivity as well as bio-compatibility.

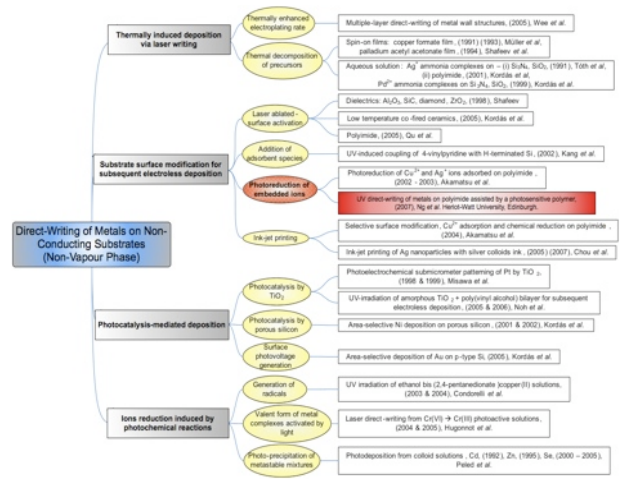

Fig. 1. Review of recent research in direct-writing of metals on non-conducting substrates. The work described in this article is highlighted.

The ability to direct-write metals on polyimide substrates in air without using the conventional photolithography method can thus motivate developments of innovative applications such as advanced interconnections, chip packaging in microelectronics, integration of electrical functions in Bio-MEMS implants and micro-fluidic devices.

## II. DIRECT-WRITING METHOD

The direct-writing approach presented here is based on the simple wet chemical pre-treatment of the polyimide substrate in order to incorporate silver ions into a thin depth of the polyimide surface layer [2]. Subsequently, a thin photosensitive polymer coating is casted on top of the ion-doped polyimide. UV irradiation exposure through this over-layer coating releases electrons which lead to reduction of the silver ions which are electrostatically immobilized within the polyimide surface. The UV irradiation can be carried out via a UV lamp with a photomask or using a UV continuous-wave (CW) laser to achieve micro-patterning. After UV exposure, washing steps using deionized water and dilute $H_2SO_4$ solution remove the polymer over-layer coating and the remaining unreacted silver ions on the polyimide substrate. Finally the polyimide sample with direct-written silver patterns is annealed at 250°C to allow for re-imidization of the modified polyimide surface and diffusion and growth of the photoreduced silver particles within the polyimide surface. Conductive copper layer is then deposited selectively on the silver surface by electroless plating. Each of the above process steps are explained as follows:

*A. Wet chemical pre-treatment*

Thin Kapton® polyimide sheets were tapped onto microscope glass slides. The ion-exchange reactions start from the KOH chemical treatment of the polyimide substrates. When a polyimide substrate is immersed into 5M KOH solution at 50°C for 5 minutes, the surface becomes hydrolyzed to a depth of at least 1 μm, in the form of a polyamic acid potassium salt. After rinsing with deionized water, the substrate is immersed in a 0.1 M AgNO₃ solution at room temperature for 30 minutes in order to exchange sufficient Ag⁺ ions from the electrolyte solution with the K⁺ ions in the substrate. KOH and AgNO₃ are both low cost, common inorganic salts. They are non-hazardous materials therefore no high waste disposal cost is incurred. The immersion processes does not require inert gas atmosphere and although a clean room environment may be preferred, it is not critical to the doping of ions into the polyimide substrates.

*B. Coating of photoreactive polymer reducing agent*

The polymer methoxy poly(ethylene glycol) (MPEG) can act as a reducing agent under UV irradiation [3-7]. This polymer is utilized as an over-layer coating to enable the direct-writing process presented here. The MPEG is dissolved in absolute ethanol with a concentration of 100 g.l⁻¹. At this concentration and using ethanol as a solvent instead of methanol or water, a smooth, thin (~1 μm) transparent coating is casted by spin-coating at 4000 rpm for 30 seconds. The MPEG solution can also be used to dip-coat contour surfaces for direct-writing.

*C. UV direct-writing*

The patterning process using a mercury arc lamp or a continuous wave (CW) UV laser can be carried out in air or in a clean room since the MPEG coating layer is solid at atmospheric pressure and does not need to be treated in any inert gas environment. This makes the handling of the coated substrates under UV exposure convenient and advanced manipulation such as 3D direct-writing can be achievable (Fig. 6). In addition, low power CW UV direct-writing has an important advantage over heat dominated laser writing processes because, in those processes, diffusion of heat can give rise to spreading, irregularity of the final deposit, damage and debris on the substrate [8]. A HeCd laser (λ = 325 nm, P = 3 mW after focus) has demonstrated successful direct-writing.

*D. Post-exposure anneal*

After washing off the MPEG coating with deionized water and exchanging the unreacted silver ions with H⁺ ions from dilute H₂SO₄ acid solution, the silver patterned polyimide is annealed at 250°C in an oven for 30 minutes. The polyimide surface layer modified by the KOH solution is re-imidized back into the original polyimide molecular structure, whilst the silver metal particles encapsulated in this polymer matrix layer migrate and grow into bigger clusters.

*E. Electroless copper plating*

The silver clusters formed on the substrate are isolated by the polyimide polymer matrix. However, these silver surfaces serve as suitable nucleation sites which can initiate electroless metal crystal growth. Therefore the direct-written silver patterns act as a catalyst seed layer for selective electroless plating. A conventional low pH formaldehyde based electroless copper solution is used which contained 16 g.l⁻¹ CuSO₄, 48 g.l⁻¹ NaK-tartrate, 28 g.l⁻¹ NaOH and 12 ml.l⁻¹ aqueous solution (37.2 wt.%) of formaldehyde. Well-defined conductive copper was electrolessly deposited exclusively onto the patterned silver seed.

## III. CHARACTERISTICS OF COPPER DEPOSITS

Greek Cross test structures (Fig. 2-5) were fabricated via an acetate photomask and a mercury arc lamp with power intensity ~ 48 mW cm⁻². A range of line widths down to 50 μm was achieved with were all in good agreement with the photomask. The deposited copper has a shinny appearance, well defined edge and good adhesion to the substrate. The sheet resistance $R_s$ of the test structure with 250 μm line width was found to be 0.453 Ω.cm⁻² as calculated by measuring the voltage drop $V$ across the adjacent arms of the cross with an applied current $I$ across the two other arms of the cross:

$$R_s = \pi/(\ln 2) \cdot V/I. \qquad (1)$$

The 3-D helix silver coil fabricated by laser direct-writing on a cylindrical polyimide substrate (Fig. 6) has continuity over the whole length of the helix (~ 28 cm) and a line width of ~ 15 μm which is in good agreement with the laser spot size employed.

## IV. CONCLUSION

A novel direct-writing fabrication process for micro-patterning of metals on polyimide is presented. This process can be carried out in air, on contoured surfaces, and requires only low fluence UV light sources and low cost source materials. The line widths of the tracks fabricated by UV-photomask method were in good agreement with that of the mask. Using laser direct-writing, smaller track width can be achieved with accuracy. Implementation of this method for fabrication of MEMS packaging and other devices requires reliable control over the resistivity, adhesion, feature size and thickness of the deposit.

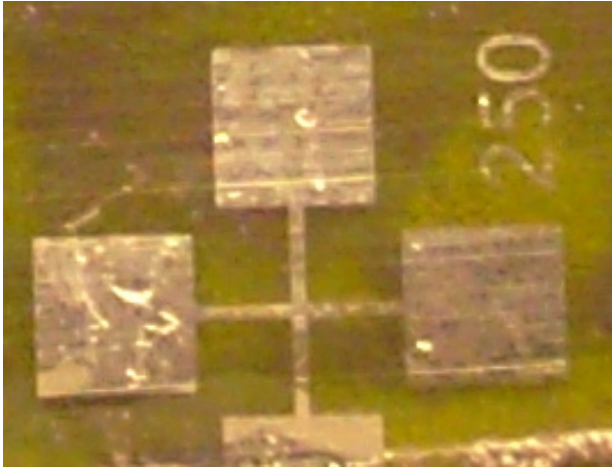

Fig.2. Greek Cross test structure with 250 µm line width.

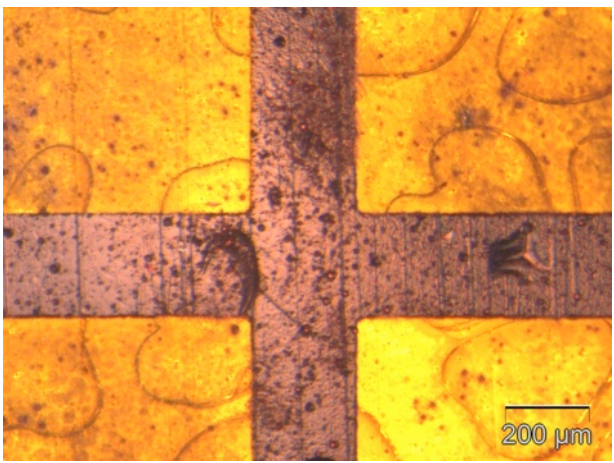

Fig.3. Magnified image of the structure shown in Fig. 1.

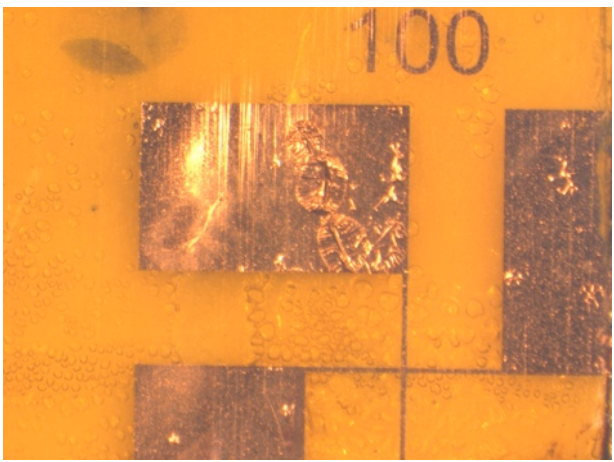

Fig.4. Test structure with 100 µm line width.

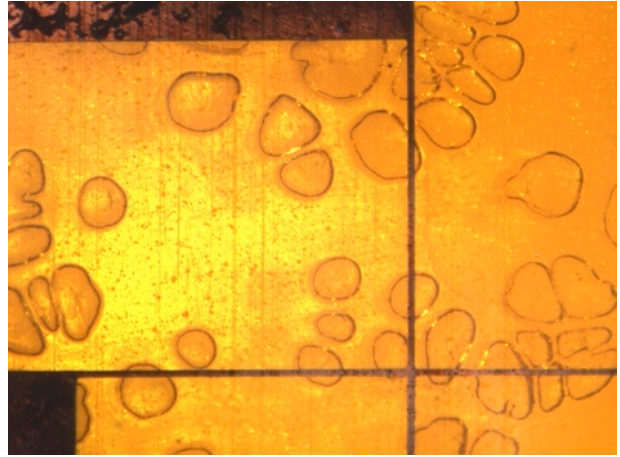

Fig.5. Test structure with 50 µm line width.

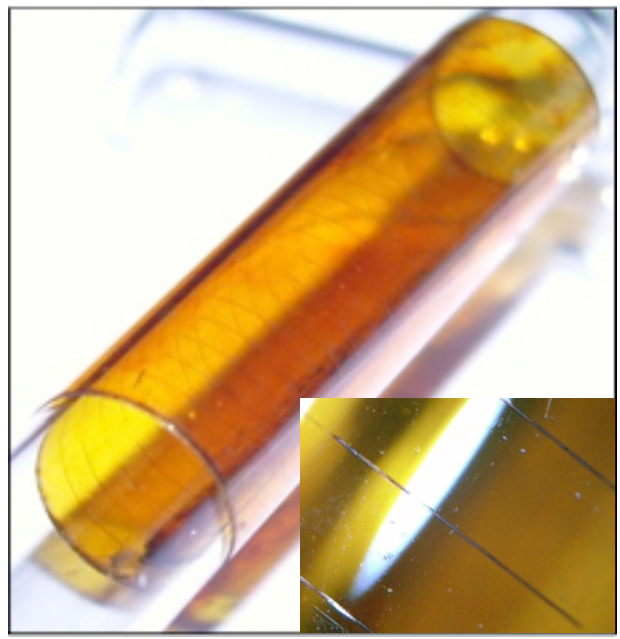

Fig. 6. Long silver helix track fabricated by rotating the cylindrical polyimide substrate (on a 1 cm diameter glass pipette) on a horizontal transition stage under the laser. Inset shows a magnified image of the track with a line width of $\sim 15\,\mu m$.

ACKNOWLEDGMENT

The authors would like to thank for the financial support from the UK Engineering Physical Sciences Research Council (EPSRC) through its Basic Technology Program. The work was conducted under the project entitled "A thousand Micro-emitters per square millimeters" referenced GR/S85764. The authors also acknowledge the support of EPSRC-funded project 3D-Mintegration (www.3d-mintegration.com).